# HIV-prevalence mapping using Small Area Estimation in Kenya, Tanzania, and Mozambique at the first sub-national level

**Enrique M Saldarriaga[1]**

[1]The Comparative Health Outcomes, Policy & Economics (CHOICE) Institute, University of Washington, Seattle, WA

**Abstract**: 221 words

**Text:** 2,076 words

**Tables & Figures:** 5

**References:** 18

**Abstract**

Local estimates of HIV-prevalence provide information that can be used to target interventions and consequently increase the efficiency of the resources. This closer-to-optimal allocation can lead to better health outcomes, including the control of the disease spread, and for more people. Producing reliable estimates at smaller geographical levels can be challenging and careful consideration of the nature of the data and the epidemiologic rational is needed. In this paper, we use the DHS data phase V to estimate HIV prevalence at the first-subnational level in Kenya, Tanzania, and Mozambique. We fit the data to a spatial random effect intrinsic conditional autoregressive (ICAR) model to smooth the outcome. We also use a sampling specification from a multistage cluster design. We found that Nyanza (P=14.2%) and Nairobi (P=7.8%) in Kenya, Iringa (P=16.2%) and Dar es Salaam (P=10.1%) in Tanzania, and Gaza (P=13.7%) and Maputo City (P=12.7%) in Mozambique are the regions with the highest prevalence of HIV, within country. Our results are based on statistically rigorous methods that allowed us to obtain an accurate visual representation of the HIV prevalence in the subset of African countries we chose. These results can help in identification and targeting of high-prevalent regions to increase the supply of healthcare services to reduce the spread of the disease and increase the health quality of people living with HIV.

**Introduction**

The most important challenge for Eastern and Southern Africa to achieve the 90-90-90 HIV targets are initiation and retention to treatment.[1] The lack of resources to provide appropriate care to all people living with HIV (PLWI) limits the proportion of people that can achieve viral suppression which in turn increases the opportunity for disease spread.[2,3] Thus, it is essential to find ways to increase resources' efficiency, in order to obtain the best possible health outcomes at the lowest investment. One way to achieve this is by improving intervention's targeting across geographical areas,[4] which requires reliable information at local level to guide policy decisions. In the context of initiation and retention to care, observing the prevalence – defined as the proportion of HIV positive people in a given region geographic, demographic and temporally defined – at a sub-national level could provide means to guide resources-allocation decisions.

The main issue with prevalence mapping is that estimations across sub-national regions are needed to obtain better information but are usually based on incomplete survey data. Therefore, if no cases are recorded in a particular region, the *empirical* prevalence would be zero, but not the *hypothetical* one, which is the one we are interested. Hence, we recognize information is incomplete (i.e. not all cases have been recorded), and the modelling probability to not belong to the total population at-risk. In addition, because no completely-at-random sampling is feasible, survey data carries selection bias in the estimations. To account for this, we conduct the prevalence mapping using Small Region Estimation (SAE).[5,6]

The aim of this study is to estimate the prevalence in a sub-set of Eastern African countries at the first sub-national level, to create information relevant for policy-decision making.

**Methods**

*Data*
We used the DHS HIV and geospatial data.[7] The selected sub-set of countries are neighbor countries in Eastern Africa, defined as sharing a border, have important differences in their national HIV prevalence estimates, and have HIV and geospatial information in the same phase of the DHS Program for sampling design and data collection consistency. We decided for: Kenya, Tanzania, and Mozambique. Whose estimated national prevalence for the 15 to 49 years old population was 5.66%, 3.97%, and 11.98%, respectively in 2017.[8,9] This is the most heterogeneous cluster, in terms of national HIV-prevalence, in East Africa that we could found. All these countries have spatial and HIV data collected in the phase DHS-V, executed between 2007 and 2009. The prevalence mapping would be conducted over the first administrative sub-national areas: regions within countries.

The geospatial information for each country was originally formed by the cluster information and the boundaries. The former included the spatial points for all clusters within regions, while the latter included the polygons for each region within a country. Using the coordinates of both datasets we identified to which region each cluster belonged. Then, we merge the information. The HIV data was geographically identified only by clusters. We used the merged spatial information to determine to which region, each of the clusters in the HIV data belong to.

The final spatial dataset is the appended data of the three countries containing the polygons for each of the sub-national regions. The final HIV dataset is also the appended data of the three countries containing the individual health outcome (HIV positive or negative), the cluster, and associated sub-national region.

*Model*

Considering that the DHS survey used a multistage cluster design sampling, we conduct a SAE, with a spatial random effect intrinsic conditional autoregressive (ICAR) model[10] – the BYM model –, for a binary outcome. The random effects model has a higher precision than the direct estimation that just considers the sampling design.[11] The BYM model allows for an indirect estimation, that allows for the use of information across regions, based on defined neighbors, to perform the smoothing. A neighbor is a region with a single or multiple shared border. We follow the 'B style' of weighting to allow places with extra neighbors to have a higher influence on the results. Since we treated the sub-set of countries as a unit, the neighbor determination is not constrained within the country. Hence, the prevalence estimation *borrows* information across countries.

The estimated prevalence for each region $P_i$, where $i$ notates each region in the first sub-national level, follows the model:[12]

$$P_i = logit(p_i)$$

$$Y_i \sim N(\theta, V_i)$$

$$\theta_i = \beta_0 + \epsilon_i + S_i$$

$$\epsilon_i \sim iid\ N(0, \sigma_\epsilon^2)$$

$$S_i \sim ICAR(\sigma_S^2)$$

Where, $p_i$ is the direct (desing-based weighted) estimate of the prevalence for the region $i$; $V_i$ is the estimated variance for each region; and $\beta_0$ the intercept. This model does not include covariates. We perform the analysis using the SUMMER package[13] in R.

**Results**

The original dataset, aggregated for the three countries had a sample size of 39,575 observation (Kenya, n=7,001; Tanzania, n=15,597; Mozambique, n=16,976). After deleting the observation containing missing data in the geographic information (no missing values in the HIV data), the final sample size was 39,258 (99.2%). Given the small amount of observations missing, we did not perform a missing-at-random analysis.

We found 45 regions in total. The sample size varies highly across these regions, ranging from 308 to 1,942 observations. (Figure 1)

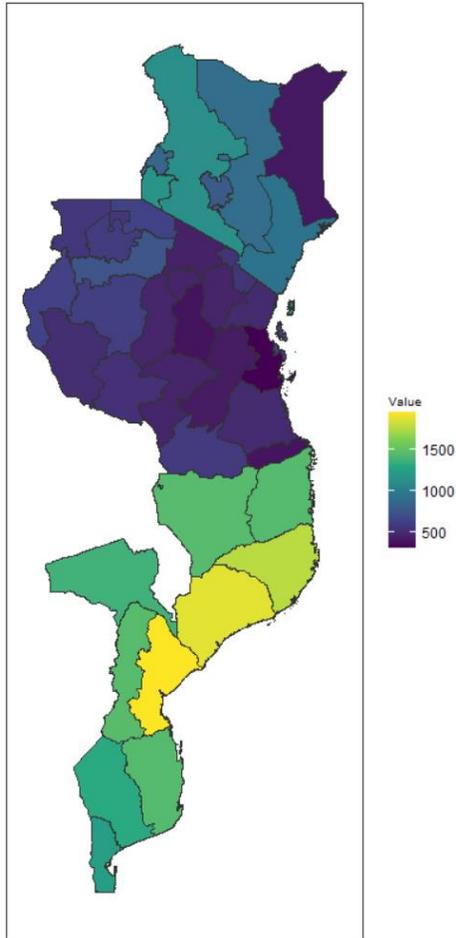

**Fig1.** Regions under analysis color-coded by their sample size

Tanzania is the country with the highest number of regions, 26, and at the same time the smallest sample size for each one. Kenya, with 8 regions, and moreover Mozambique, with 11 regions, have higher associated sample size.

Figure 2 shows the results of the SAE. The estimated prevalence for all regions ranges from 7.24% to 15.5%; with a median of 5.13% and a standard deviation (SD) of 0.38. Figure 3 shows the point-estimate results for each country. The scale for the color-code is independent for each country in Figure 3.

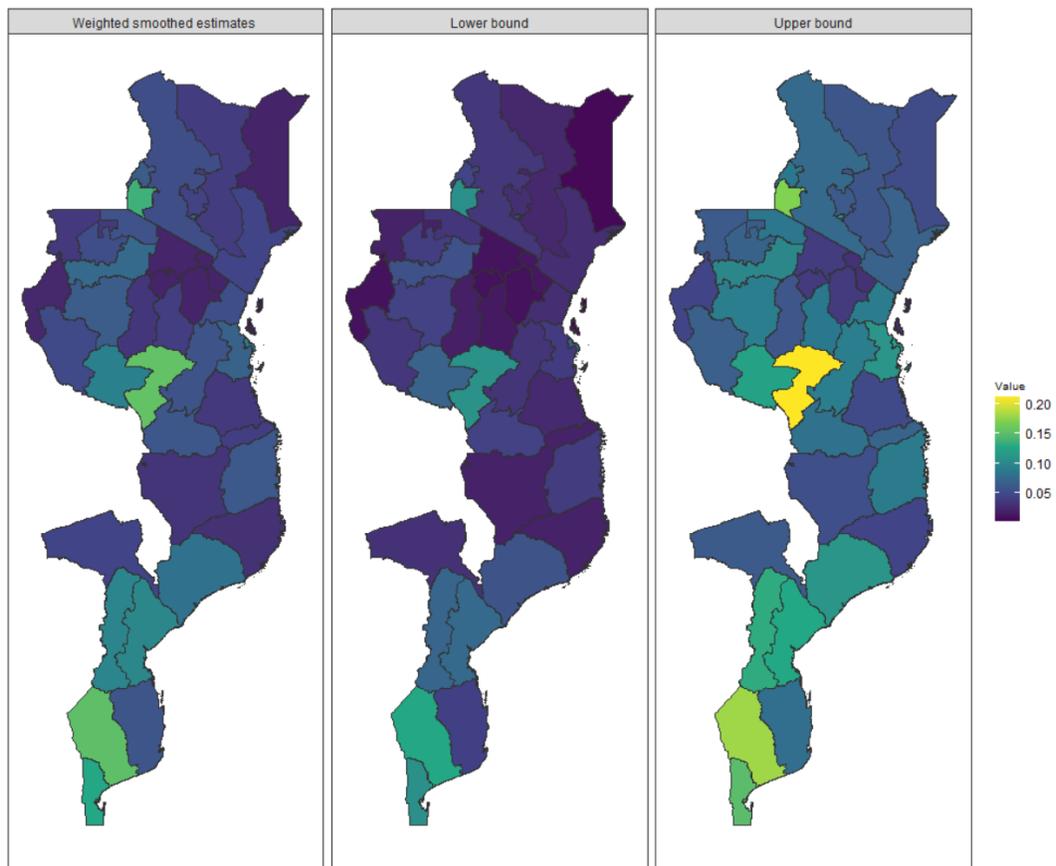

**Fig 2.** Maps of point estimate and 95% credible intervals for the posterior mean of the weighted smoothed model

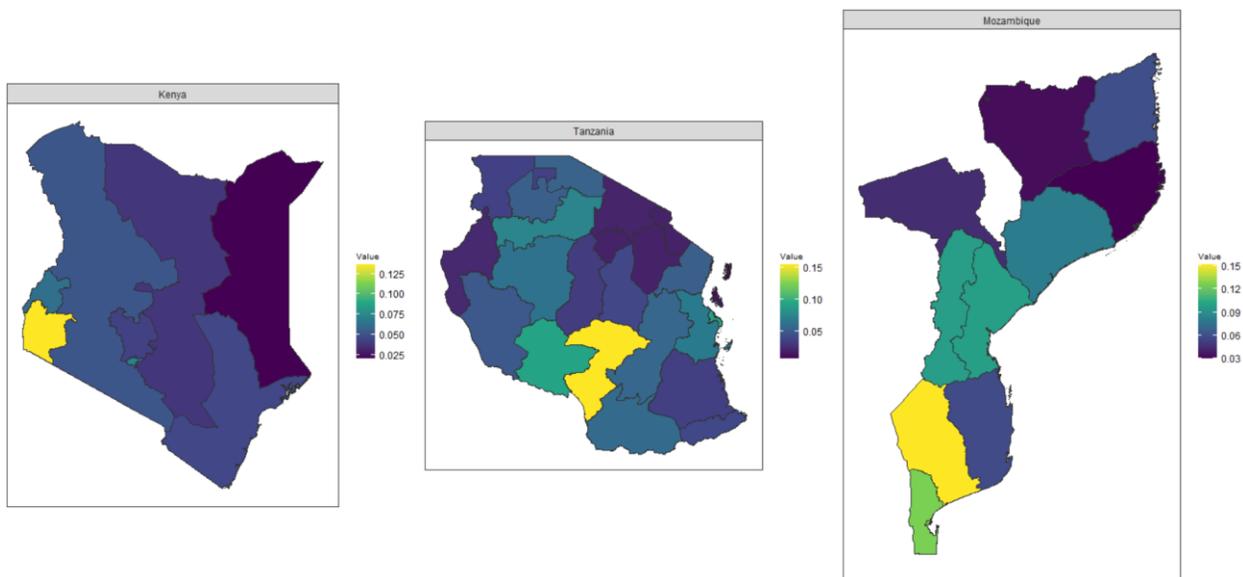

**Fig 3.** Zoom in the HIV-prevalence maps for each country. Kenya in the left, Tanzania in the middle, and Mozambique in the right panel

Our analysis allows us to identify the three regions with the highest prevalence in each country. Nyanza ($P_i$=14.2%), Nairobi ($P_i$=7.8%), and Western ($P_i$=6.4%), in Kenya. In Tanzania, Iringa ($P_i$=16.2%), Dar es Salaam ($P_i$=10.1%), and Mbeya ($P_i$=9.0%). Finally, in Mozambique, Gaza ($P_i$=14.9%), Maputo City ($P_i$=13.7%), and Maputo (region) ($P_i$=12.7%). In Kenya and Tanzania, we can observe that the highest prevalence is almost an outlier compared to the rest of the country. These results would warrant a prioritization of these regions at a Federal level.

**Discussion**

We conducted a HIV-prevalence mapping for a subset of east African countries, with a SAE estimation using a BYM model. This approximation has multiple theoretical benefits. First, the SAE is consistent with the small sample size that observed in many of the areas under estimation. Second, our model acknowledges the sampling design and the associated distribution of sampling probabilities. Third, conducts a spatial smoothing process that allows the model to borrow information from the neighbors in the estimation of each posterior to improve the precision of the estimates, which is particularly helpful in presence of sparse data as the DHS'. Our study has an additional advantage because it defines the neighbors beyond the country borders, by taking the sub-set of countries as a unit. We believe this was particularly useful in the estimation of the prevalence for Tanzania, where all its regions had a small sample size.

*Comparative performance of the smoothing process*

We present the difference in performance for the estimation of prevalence at the first-subnational level, for the weighted model that includes a correction for cluster sampling, and the weighted-smooth model that includes a smoothing process for the outcome.

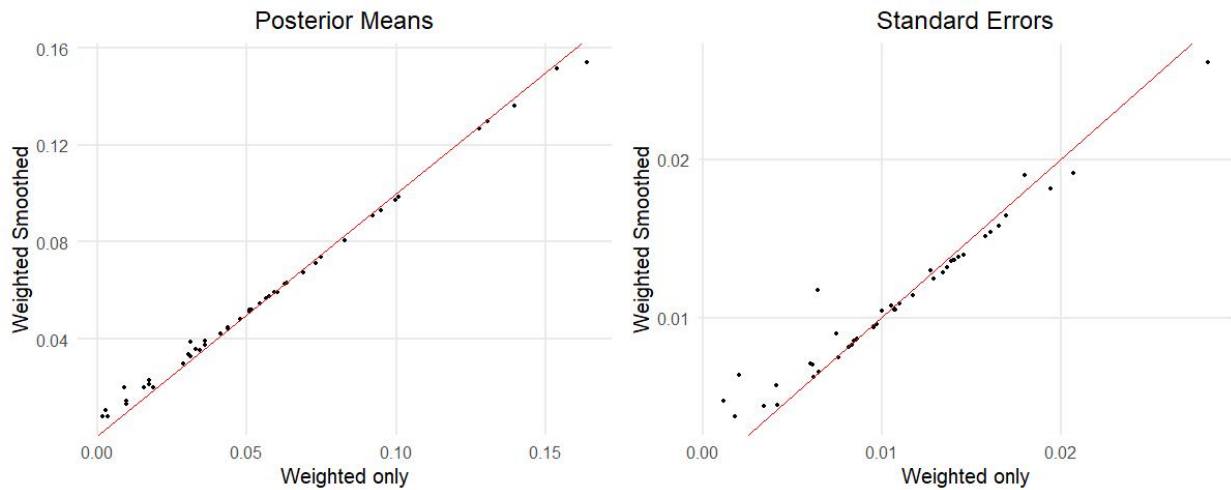

**Fig 4.** Comparison of estimated prevalence (left panel) and standard errors (right panel) between the weighted only and weighted and smoothed (SAE) estimates.

Figure 4 shows that the smoothing process pulled the posterior estimates towards the middle of the distribution, increasing the value of the estimates in the lower half, and reducing it for the estimates in the higher half. The right-side panel shows that the standard errors created by the smoothing analysis tend to be lower in comparison, denoting estimates with higher precision.

Thus, because the smoothing model estimates each prevalence using the information provided by its neighbors, the resulting distribution of prevalence for all regions is less disperse, with the extreme values pulled towards the middle. Regarding the neighbors' definition, there are two major style of weighting "B" and "W". We perform separate analyses with both styles obtaining the exact same results for the posteriors and the standard deviation. We decided in favor of the B style because in theory it allows for the areas with more neighbors to have more influence in the results. Which makes sense when considering that migration is a very important factor in the HIV epidemic[14], and regions with more borders are more likely to have more migrants.

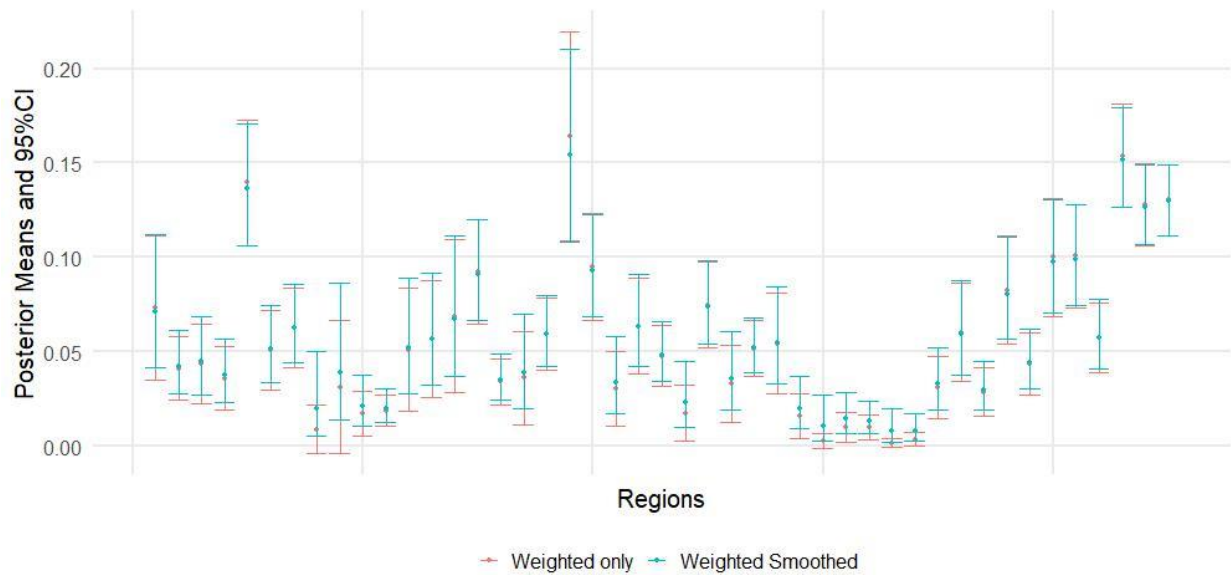

**Fig 5.** Error bars for the Weighted smoothed (SAE) estimates and the Naïve (direct) ones

According to Figure 5, even though the weighted only model includes a correction for the sampling methodology, the resulting estimates for around 4 regions are still not significantly different from zero i.e. estimations whose lower 95%CI bound included zero. This demonstrates an important consequence of the smoothing process, the elimination of prevalence non-statically different from zero. This issue weights on the difference between the observed and the empirical prevalence.[6] The lack of cases in a given cluster does not mean that there are no cases in that geographical area. Hence, surrogating the estimated risk from adjacent areas is imperative to estimate a prevalence closer to the *real* risk of infection for HIV.

Compared to previous analysis that have mapped the prevalence of HIV[15], our results are systematically lower. One reason resides in the granularity of the data used by the authors. They collected 38,897 data points in 134 seroprevalence survey and sentinel surveillance of antenatal care clinics for a total of 46 countries. Moreover, they gathered information for 17 years. Hence, the analytical process can smooth the estimates across geographic and temporal units and provide better estimates. Additionally, the authors calibrated the model using national estimations from the Global Burden of Disease[16], which provides an additional level of external validity but also introduces more bias in the analysis. Nonetheless, the estimated confidence intervals for both studies are overlapped in all the cases for which a comparison was possible, and more importantly our results are consistent regarding which regions within countries have the highest prevalence.

### *Limitations*

This study is not without limitations. First, the data that we used was collected between 2007 and 2009. Hence, the picture that our estimates present is likely outdated considering how fast the HIV epidemic variables move over time. We chose the phase DHS-V because that was the only phase in which our sub-set of countries had HIV and geospatial information.

Second, although socioeconomic and demographic variables influence the prevalence of HIV[17,18] and this variables are likely spatial correlated, our model did not include covariates. The main reason is that we did not find a linkage between the HIV data and the individual survey from the DHS as well, that contains many of those variables. Further investigation is necessary.

### *Strengths*

The objective of this study was to present with reliable information to inform policy decision in the context of optimize resources allocation. In that regard, the temporality of the data is not as important since the absolute numbers can change by the relative risk is less likely to change or even shift over time.

Second, our methodology is in concordance with both the nature of the data – by including a correction due to the sampling design – and the epidemiology rationale – by using a smoothing process to limit the probability of having prevalence equal to zero and modeling the outcome in each region as part of a bigger area, rather than in isolation.

### *Conclusions*

Our results are based on statistically rigorous methods that allowed us to obtain an accurate visual representation of the HIV prevalence in the subset of African countries we chose. These results can help in identification and targeting of high-prevalent regions to increase the supply of healthcare services to reduce the spread of the disease and increase the health quality of PLWH.

This study builds on secondary publicly available data and generates reliable estimates that can be used to target interventions. This creates an important opportunity to apply the same methodology in other settings, where data might be scarce and resources to supplement data collection unavailable.

**References**

1. Marsh, K. *et al.* Global, regional and country-level 90–90–90 estimates for 2018: assessing progress towards the 2020 target. *AIDS* **33**, S213 (2019).
2. Kay, E. S., Batey, D. S. & Mugavero, M. J. The HIV treatment cascade and care continuum: updates, goals, and recommendations for the future. *AIDS Res. Ther.* **13**, (2016).
3. Kharrazi, H., Chang, H.-Y., Heins, S. E., Weiner, J. P. & Gudzune, K. A. Assessing the Impact of Body Mass Index Information on the Performance of Risk Adjustment Models in Predicting Health Care Costs and Utilization. *Med. Care* **56**, 1042–1050 (2018).
4. McGillen, J. B., Anderson, S.-J., Dybul, M. R. & Hallett, T. B. Optimum resource allocation to reduce HIV incidence across sub-Saharan Africa: a mathematical modelling study. *Lancet HIV* **3**, e441–e448 (2016).
5. Rao, J. N. K. & Molina, I. *Small area estimation*. (John Wiley & Sons, Inc, 2015).
6. Pfeffermann, D. New Important Developments in Small Area Estimation. *Stat. Sci.* **28**, 40–68 (2013).
7. ICF. The DHS Program - Available Datasets. Funded by USAID. https://dhsprogram.com/data/available-datasets.cfm.
8. Frank, T. D. *et al.* Global, regional, and national incidence, prevalence, and mortality of HIV, 1980–2017, and forecasts to 2030, for 195 countries and territories: a systematic analysis for the Global Burden of Diseases, Injuries, and Risk Factors Study 2017. *Lancet HIV* **6**, e831–e859 (2019).
9. Roser, M. & Ritchie, H. HIV / AIDS. *Our World Data* (2020).
10. Besag, J., York, J. & Mollié, A. Bayesian image restoration, with two applications in spatial statistics. *Ann. Inst. Stat. Math.* **43**, 1–20 (1991).
11. Fay, R. E. & Herriot, R. A. Estimates of Income for Small Places: An Application of James-Stein Procedures to Census Data. *J. Am. Stat. Assoc.* **74**, 269–277 (1979).
12. Wakefield, J. Prevalence Mapping. *Wiley StatsRef* 1–7 (2019) doi:10.1002/ISBN.stat00999.pub9.
13. Martin, B. *et al. SUMMER: Spatio-Temporal Under-Five Mortality Methods for Estimation. R package version 0.2.1.* (R Foundation for Statistical Computing, 2018).
14. Nöstlinger, C. & Loos, J. Migration patterns and HIV prevention in Uganda. *Lancet HIV* **5**, e158–e160 (2018).
15. Dwyer-Lindgren, L. *et al.* Mapping HIV prevalence in sub-Saharan Africa between 2000 and 2017. *Nature* **570**, 189–193 (2019).
16. James, S. L. *et al.* Global, regional, and national incidence, prevalence, and years lived with disability for 354 diseases and injuries for 195 countries and territories, 1990–2017: a systematic analysis for the Global Burden of Disease Study 2017. *The Lancet* **392**, 1789–1858 (2018).
17. Edwards, A. E. & Collins, C. B. Exploring the influence of social determinants on HIV risk behaviors and the potential application of structural interventions to prevent HIV in women. *J. Health Disparities Res. Pract.* **7**, 141–155 (2014).
18. Chanda-Kapata, P., Klinkenberg, E., Maddox, N., Ngosa, W. & Kapata, N. The prevalence and socio-economic determinants of HIV among teenagers aged 15–18 years who were participating in a mobile testing population based survey in 2013–2014 in Zambia. *BMC Public Health* **16**, 789 (2016).